\documentclass{PoS}

\title{Slavnov-Taylor Identities in Coulomb Gauge Yang-Mills Theory}

\ShortTitle{Slavnov-Taylor Identities in Coulomb Gauge Yang-Mills Theory}

\author{\speaker{Peter Watson}%
        \thanks{Work supported by the Deutsche Forschungsgemeinschaft (DFG) 
        under contracts no. DFG-Re856/6-1 and DFG-Re856/6-2.}\\
        T\"ubingen University\\
        E-mail: \email{watson@tphys.physik.uni-tuebingen.de}}

\author{Hugo Reinhardt\\
        T\"ubingen University\\}
\abstract{
Two aspects of the color charge in Coulomb gauge continuum Yang--Mills 
theory are discussed.  The first aspect is the existence of a conserved and 
vanishing total charge exhibited within the first order functional 
formalism.  The second aspect is the closure of the set of Slavnov--Taylor 
identities in the second order functional formalism, such that the exact 
solution for temporal Green's functions is in principle possible and thereby 
preserving the color charge.
}

\FullConference{8th Conference Quark Confinement and the Hadron Spectrum\\
                 September 1-6, 2008\\
                 Mainz. Germany}

\begin{document}
%=============================================================================

Coulomb gauge offers great potential in understanding confinement.  There are, however, several outstanding issues.  Firstly, the inherent noncovariance makes even perturbative integrals difficult to compute (but these are being done \cite{Watson:2007mz,Watson:2007vc}).  Second, there exist potential energy divergences -- ghost loops without energy dependent components exhibit a pure and unregularizable divergence; where analyzed though, it is seen explicitly that such loops cancel.  Third is the issue of incomplete gauge-fixing; after fixing to Coulomb gauge, purely temporal gauge transforms are still allowed.  Fourth is the Gribov problem, where it is recognized that the standard Faddeev-Popov method does not completely fix the gauge \cite{Gribov:1977wm}.

We will (partly) address the latter three issues in the first part of this presentation by considering the temporal zero-modes of the Faddeev--Popov operator and how a conserved, vanishing total charge for the gluon field arises \cite{Reinhardt:2008pr}.  In the second part, we look at the Coulomb gauge Slavnov--Taylor identities (STids); these are the dynamical implementation of charge conservation \cite{Watson:2008fb}.

%=============================================================================
Let us start with the temporal zero-modes.  We consider continuum Yang--Mills theory and the generic functional integral (following the conventions of Ref.~\cite{Watson:2006yq})
\begin{equation}
Z=\int{\cal D}\Phi\exp{\left\{\imath{\cal S}\right\}},\;\;\;\;
{\cal S}=\frac{1}{2}\int d^4x\left[E^2-B^2\right]
\label{eq:origact}
\end{equation}
where the chromoelectric field, $\vec{E}^a$, is written in terms of spatial ($\vec{A}$) and temporal ($\sigma$) gauge-field components:
\[
\vec{E}^a=-\partial^0\vec{A}^a-\vec{\nabla}\sigma^a+gf^{abc}\vec{A}^b\sigma^c=-\partial^0\vec{A}^a-\vec{D}^{ac}[\vec{A}]\sigma^c
\]
(the chromomagnetic field, $\vec{B}^a$, is a function of $\vec{A}$ only).  The action ${\cal S}$ is invariant under gauge transforms parametrized by $\theta$ and we choose Coulomb gauge ($\vec{\nabla}\cdot\vec{A}=0$), using the Faddeev--Popov method
\[
1=\int{\cal D}\theta\delta\left(\vec{\nabla}\cdot\vec{A}^a\right)\mbox{Det}\left[-\vec{\nabla}\cdot\vec{D}\right].
\]
However, when $\theta$ is spatially independent, the determinant vanishes and there are temporal zero-modes.  We replace $\mbox{Det}$ in the identity above with $\overline{\mbox{Det}}$, the determinant with the zero-modes removed; correspondingly, $\theta$ is also restricted.  We convert to the first order formalism by introducing an auxiliary field $\vec{\pi}$ and splitting this into transverse ($\vec{\nabla}\cdot\vec{\pi}^{\bot a}=0$) and longitudinal ($\vec{\nabla}\phi$) parts \cite{Zwanziger:1998ez,Watson:2006yq}.  In doing this, the action becomes linear in $\sigma$ so we can integrate it out to leave a $\delta$-functional constraint on the functional integral:
\[
Z=\int{\cal D}\Phi\delta\left(\vec{\nabla}\cdot\vec{A}^a\right)\delta\left(\vec{\nabla}\cdot\vec{\pi}^{\bot a}\right)\overline{\mbox{Det}}\left[-\vec{\nabla}\cdot\vec{D}\right]\delta\left(\vec{\nabla}\cdot\vec{D}^{ab}\phi^b+g\hat{\rho}^{a}\right)\exp{\left\{\imath{\cal S}'\right\}}
\]
where $\hat{\rho}^{a}=f^{abc}\vec{A}^b\cdot\vec{\pi}^{\bot c}$ is the color charge of the gauge field.  Due to the $\delta$-functional constraint, the $\phi$-field can also be integrated out, \emph{but}, the zero-modes of the operator $-\vec{\nabla}\cdot\vec{D}$ must be taken properly into account.  Doing this, we arrive at the form \cite{Reinhardt:2008pr}
\[
Z=\int{\cal D}\Phi\delta\left(\vec{\nabla}\cdot\vec{A}^a\right)\delta\left(\vec{\nabla}\cdot\vec{\pi}^{\bot a}\right)\delta\left(\int d^3x\hat{\rho}_x^a\right)\exp{\left\{\imath{\cal S}''\right\}}
\]
where the modified determinant cancels out and with the final effective action (no longer invariant under temporal gauge transforms since $\sigma$ has been integrated out)
\[
{\cal S}''=\int d^4x\left[\vec{\pi}^{\bot}\cdot\partial^0\vec{A}-\frac{\pi^{\bot 2}}{2}-\frac{B^2}{2}+\frac{g^2}{2}\hat{\rho}\left(-\vec{\nabla}\cdot\vec{D}\right)^{-1}\nabla^2\left(-\vec{\nabla}\cdot\vec{D}\right)^{-1}\hat{\rho}\right].
\]
What we see is that the temporal zero-modes cancel out (i.e., the incomplete gauge-fixing is not a problem), leaving behind a vanishing, conserved total charge ($\int d^3x\hat{\rho}_x^a$) and only physically transverse degrees of freedom with a gauge-fixed action.  The (modified) Faddeev--Popov determinant cancels so there are also no more energy divergences.

%=============================================================================
To discuss the STids, let us return to the original action, Eq.~(\ref{eq:origact}).  We fix to Coulomb gauge and introduce ghosts in the usual way.  The action is invariant under Gauss-BRST transforms -- standard BRS transforms, but with a \emph{time-dependent} Grassmann variation, $\delta\lambda_t$, since the ghost terms only involve spatial derivatives \cite{Zwanziger:1998ez} (usually $\delta\lambda$ is global).  After Fourier transform, the time-dependence of $\delta\lambda_t$ serves to inject an additional energy scale ($q_0$) into the identities.  The STids arise from considering the gauge transformation as a change in variables within the functional integral and taking various functional derivatives to obtain expressions for the Green's functions.  The STids are the local expression of charge conservation.

For the proper two-point Green's functions, we have identities of the form \cite{Watson:2008fb}
\[
k_0\Gamma_{\sigma\sigma}^{ab}(k)=\imath\frac{k_i}{\vec{k}^2}\Gamma_{\sigma Ai}^{ac}(k)\Gamma_{\overline{c}c}^{cb}(q_0+k),\;\;\;\;
k_0\Gamma_{A\sigma j}^{ab}(k)=\imath\frac{k_i}{\vec{k}^2}\Gamma_{AAji}^{ac}(k)\Gamma_{\overline{c}c}^{cb}(q_0+k)
\]
where $\Gamma_{\overline{c}c}$ is the proper ghost function.  All of the temporal (external $\sigma$-leg) Green's functions are given in terms of spatial ($\vec{A}$-leg) and ghost functions.  Since $q_0$ is arbitrary, the ghost functions must be independent of energy.  These identities have been checked at one-loop perturbatively \cite{Watson:2007vc}.  For the three-point proper functions we have a series of identities of the type \cite{Watson:2008fb}
\[
k_3^0\Gamma_{XY\sigma}=\imath\frac{k_{3i}}{\vec{k}_3^2}\Gamma_{XYAi}\Gamma_{\overline{c}c}+\tilde{\Gamma}
\]
where $\tilde{\Gamma}$ represents a series of ghost kernels and two-point functions.  These kernels also involve temporal Green's functions, but Coulomb gauge does something rather special.  When we take further functional derivatives to get higher order STids three things happen: 1) the arbitrary energy scale ($q_0$) removes any symmetry when swapping legs, allowing for a unique `left-hand side', 2) temporal legs are scalar, so there are no undetermined `transverse' parts (unlike covariant gauges) and 3) ghosts come in pairs -- adding such a pair restricts the possibilities for temporal factors in the loops that make up the kernels.  Eventually, we find that the STids for the three-point functions are part of a closed set of seven equations \cite{Watson:2008fb}.  Given the spatial and ghost functions as input, the temporal Green's functions are the solution.  This means that charge conservation can be enforced locally and nonperturbatively (in principle, exactly).

%=============================================================================

%=============================================================================

%=============================================================================
\end{document}